\documentclass[aps,prb,showpacs,showkeys,amssymb,amsmath,superscriptaddress]{revtex4}

\usepackage{graphicx}
\usepackage{dcolumn}
\usepackage{bm}
\usepackage[version=3]{mhchem}
\usepackage{float}
\usepackage[english]{babel}
\usepackage{color}
\usepackage{tabularx}

\begin{document}


\title{
\ce{FeAs2} formation and electronic nematic ordering:
an analysis in terms of structural transformations
}

\author{A. Pishtshev}%
 \email{aleksandr.pishtshev@ut.ee}
  \affiliation{%
Institute of Physics, University of Tartu, Ravila 14c, 50411 Tartu, Estonia\\
          }%

\author{P. Rubin}%
  \affiliation{%
Institute of Physics, University of Tartu, Ravila 14c, 50411 Tartu, Estonia\\
          }%

\date{\today}

\begin{abstract}
By combining DFT-based computational analysis and symmetry constraints in terms of group-subgroup relations,
we analysed the formation of the native crystalline structure of loellingite \ce{FeAs2}.
We showed that the ground state of the material exhibits the ordered patterns of the  electronic localization which are mainly associated with iron $3d_{ x^2 - y^2}$ orbitals and can be characterized in terms of nematic-like ordering.
The ordering is the result of the close interplay of the lattice and the electron degrees of freedom.
In a structural aspect, it pursues an energy quest to select the orthorhombic crystal lattice attributed to the $Pnnm$ space group.
In a charge aspect, the ordering is connected with the valence charge density redistribution that not only provides a high electronic polarizability but also  gives rise to an extra-large magnitude of the negative component of the dynamical p-d charge transfer.
\end{abstract}




\maketitle


\section{Introduction}
The structural integrity of a bulk solid containing a transition metal (TM) element
with an open $3d$-shell is governed by a mutual balance of the driving chemical interactions.
In this aspect, the direct charge transfer, the overlap of the outer orbitals, and the many-body effects of electronic correlations appear to be the most significant structure-dependent factors.
On the other hand, the cooperative character of chemical bonding in the periodic lattice
determines the activity of these interactions upon a wide range of length-scales.
The effects of electronic correlations plays a very important role because they tend to keep
an atomic format of partially filled $3d$ states.
This factor causes charge localization on short-range scales and thereby gives extra opportunities
for valence electrons to affect the underlying chemical structure, crystal packing geometry,
and electron sharing possibilities.

Thus, the flexible theoretical characterization of the crystal design of a compound with a $3d$ TM element should greatly depend on an understanding of how the typical properties of structural geometry of the solid state may be influenced by the features of spatial localization patterns of valence electrons.
In the article, using the pnictide \ce{FeAs2} as a model example of a binary TM-nonmetal alloy
we ascertain how the interplay between lattice stabilization and valence electron localization
sets up interesting peculiarities of the crystal chemistry of this material.
To gain full insight, we investigated the electronic properties caused by the $3d$ electrons of
\ce{Fe} cations in terms of possible structural transformations driven by a TM-ligand coordination. Based on the obtained results, we found that the formation of a preferred architecture of \ce{FeAs2}
is accompanied by the further utilization of \ce{Fe-Fe} planar interactions with the establishment of nematic-like ordering.
\section{Material, methods and simulation details}
Loellingite belongs to a family of binary compounds (called  pnictides) that contain
 a combination of a TM with the ligands consisting of Group 15 elements
such as \ce{P}, \ce{As}, \ce{Sb}, and \ce{Bi}.\cite{R_FeAs2}
The orthorhombic structure of \ce{FeAs2} is characterized by
the space group $Pnnm$  (compressed marcasite phase) with lattice parameters
a=5.3012 {\AA}, b=5.9858 {\AA}, c=2.8822 {\AA}.\cite{2}
A study of the \ce{Fe-As} phase diagram showed \cite{2a} that  \ce{FeAs2} retains stability
up to its melting point at 1020$^\circ$ C.
The insulating nature of FeAs$_2$ was observed in resistance measurements \cite{vapour} performed for
a single crystal at ambient temperatures. An estimate of the band gap made from
an analysis of the experimental resistivity $\rho$
behavior {\it vs.} $T$ in terms of $log \rho$ indicated a value of about $0.22$ eV.
Computational studies of the electronic structure of \ce{FeAs2} provided information
on the main features of electron bonds.\cite{3}

To gain a deeper insight into the characteristic details
of $3d$ iron (and $4p$ arsenic orbitals) and to understand the structural factors
markedly affecting the electron localization in \ce{FeAs2}, we analyzed the ground state
by employing DFT+U theory\cite{4} and the G$_0$W$_0$ approximation\cite{5}, with an emphasis on determining
the role the coordination environment plays in the genesis of the electronic subsystem.
We performed spin-polarized periodic calculations using the Vienna {\it ab initio} simulation
package (VASP)\cite{6} together with the potential projector augmented-wave (PAW) method.\cite{7}
As iron belongs to the category of first-row late transition metal elements with localizable $d$ orbitals,
we determined the Kohn-Sham (KS) eigenstates within the Perdew-Burke-Ernzerhof (PBE) GGA exchange-correlation
functional\cite{8} augmented with the Hubbard U term.
The approach of Dudarev {\it et al}.\cite{9} has been used for PBE+U calculations.
The effective value of the on-site $d$-$d$ Coulomb repulsion, which represents the difference of
the intra-atomic ($U$) and exchange ($J$) contributions $U_{eff}=U-J$, was estimated
via constrained PBE calculations based on linear response theory.\cite{10}
We found that an average $U_{eff}$ of $1.6$ eV can be applied to take into account electron correlations
in the description of the \ce{FeAs2} ground state. 
In accordance with experiment,\cite{2} our calculations reproduced a nonmagnetic ground state.
The topological features of the valence electron distributions
were investigated on the base of the theoretical charge densities;
a grid-based Bader analysis post-processing method,\cite{11} which implements the AIM ("atoms in molecules")
approach\cite{12} and the electron localization function (ELF)\cite{13} have been employed.
The many-body polarization effects were characterized in terms of the Born dynamical charges;\cite{14}
these charges were calculated within the scheme of density functional perturbation theory.\cite{15}
Analysis of the structural properties was performed by using the programs FINDSYM of the ISOTROPY Software Suite\cite{16}
and VESTA.\cite{17} For visualizations of the structures and electron topologies, we applied the VESTA program.
Group-subgroup sequences forming the B{\"a}rnighausen tree\cite{18} were found by using the programs
INDEX and SUBGROUPGRAPH\cite{19} hosted by the Bilbao Crystallographic Server.\cite{20,21}

A plane-wave basis set with a $500$ eV cutoff and a $\Gamma$ point centered mesh for the ${\bf k}$-point sampling
have been chosen in our calculations.
For cell and atom relaxations and stability analysis, k-grid sampling was taken as $6\times6\times8$. 
$G_{0}W_{0}$ calculations
employed a $6\times6\times10$ k-point grid.
The details of the electronic structure were investigated at a denser $8\times8\times12$ k-point grid. 
The tetrahedron method with Bl{\"o}chl corrections\cite{Bc} was used for Brillouin zone integrations.
Zone centered vibrational modes and elastic moduli were obtained within the frameworks of
the finite differences method, as implemented in VASP.
\section{Results and discussion}
\subsection{Chemical content of structural formation of \ce{FeAs2} }
Although chemical considerations relate FeAs$_2$ to intermetallic compounds/metalloids (like NiAs$_2$),
the material is qualified as a low-gap semiconductor.
Our DFT calculations have shown that the relatively small Hubbard energy (U$_{eff} \sim$ 1.6 eV)
related to iron 3d electrons does not favor the metal-dielectric transition.
Note also that  both elements of the composition \ce{FeAs2} possess approximately similar values of
the Pauling electronegativity ($1.83$ for \ce{Fe} and $2.18$ for \ce{As}).
In this case, the mechanism of TM-ligand coordination, lattice structure stabilization and formation of the band gap appears an open problem
when viewed from the point of view of crystal design.
To reveal the key role the coordination environment plays in the genesis of the electronic
structure, we considered the theoretical route of the \ce{FeAs2} structural assembly.
Starting from the chemical composition, we presumed that
an assembly route of \ce{FeAs2}, written in terms of the chemical building blocks as
$$
\cee{Fe + As-As -> FeAs2},
$$
represents {\it a de novo} process under which an array of \ce{As} dimers
is packed into the iron framework to arrange  \ce{Fe-As} coordination.
In a formal chemical context, the left part of this reaction corresponds to the starting mixture of the reactants,
which is used in practical work, for example, in the chemical vapor transport method
to grow \ce{FeAs2} single crystals.\cite{vapour}
In our approach, we modeled a couple of \ce{As} atoms in terms of an intermediate dimer
to be placed in the parent \ce{Fe} lattice.
We also assumed that the structural transformation passes the chemical content
through the affinity property of rectangular lattices.
This provides an optimization of atom sites via the regular scalability of the unit-cell shape/volume
coupled with the recasting of arsenic positions in the coordination environment.

\begin{figure}[H]
\centering
\includegraphics[width=0.48\columnwidth,keepaspectratio=true]{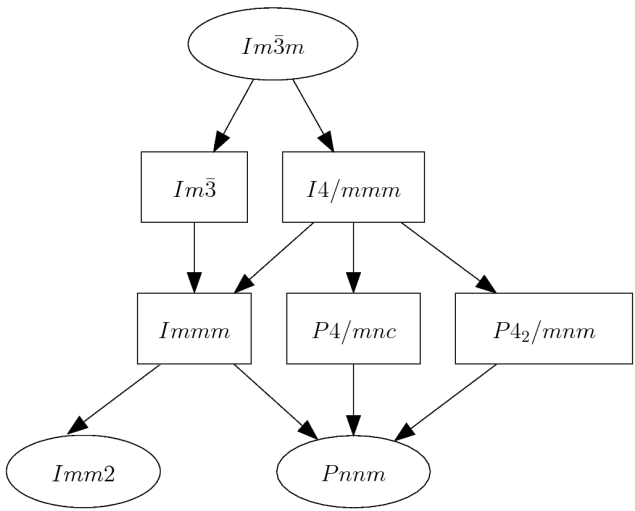}
\caption{\label{fig:Fig_1}
The B{\"a}rnighausen tree related to the loellingite structure. }
\end{figure}

\begin{figure}[H]
\centering
\includegraphics[width=0.5\columnwidth,keepaspectratio=true]{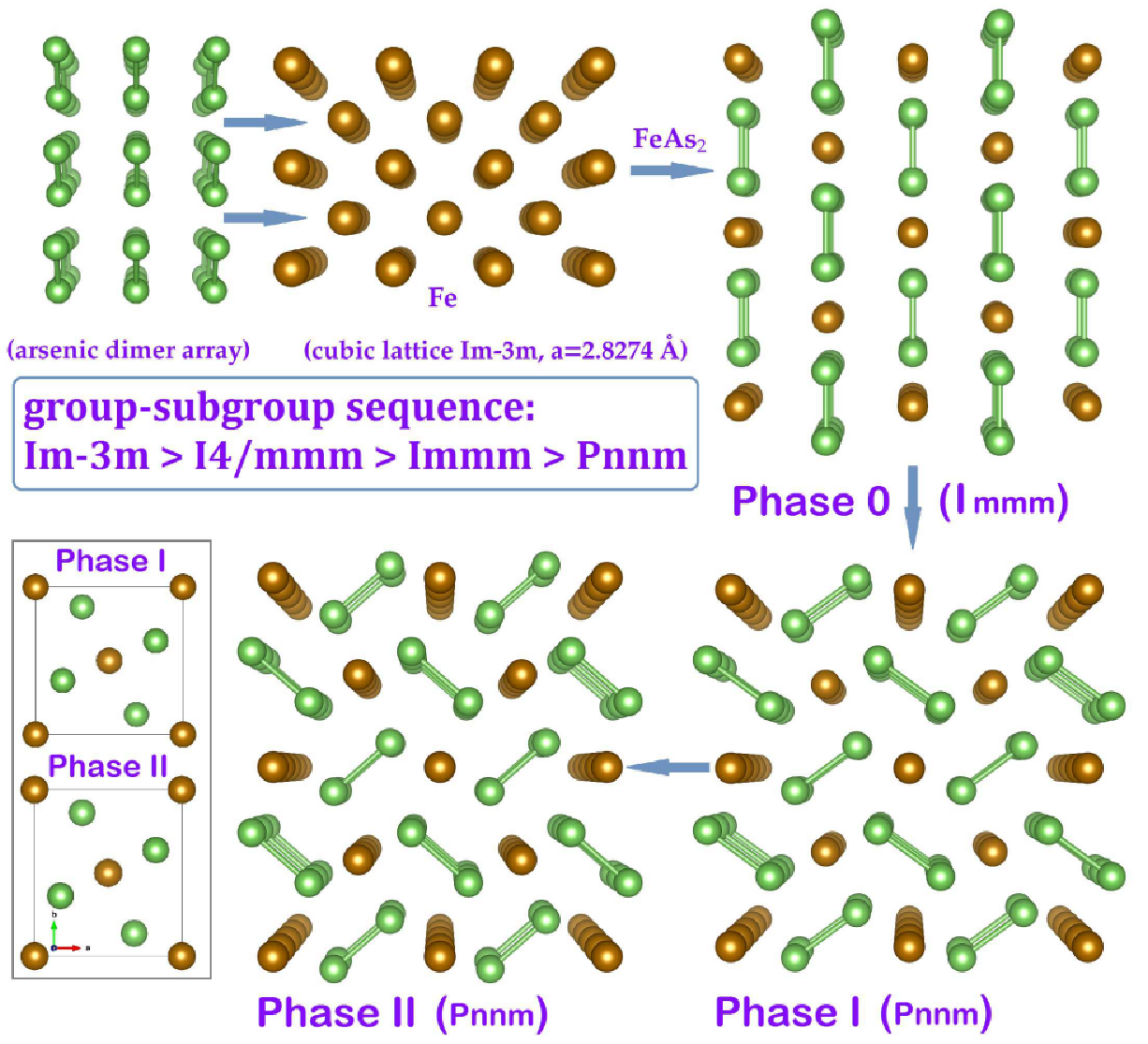}
\caption{\label{fig:Fig_2}
Flowchart of the  \ce{FeAs2} three-phase assembly between the template and the native structures.
Atoms are shown as brown (\ce{Fe}) and green (\ce{As}).
The crystallographic data are given in Appendix (Table~\ref{tab:Table_data_01}).
The packing of Phases 0, I, and II is shown as a perspective $c$-axis plot.
The common motif of  all phases is a full matching of the Wyckoff positions ('$2a$' sites)
occupied by iron atoms.
The assembly utilizes the arsenic dimer array, which is disposed along the $c$-axis by fixation to
the iron $z$-coordinate after dissolving in the iron framework of the $I4/mmm$ symmetry.
The dimer-iron linkage occurs along one of the lattice sides as the "axial alignment" of \ce{As-As} bonds.
This determines the direction of the $c$-axis because of the $O_{h} \rightarrow D_{2h}$ reduction
of the parent $Im\bar{3}m$ crystal symmetry and preassigns the smallest side, $2.8747$ {\AA}, of the unit cell.
The differences of the total energies between Phases 0 and I, and between Phases I and II are
$448$ kJ/mol and $54$ kJ/mol, respectively.
}
\end{figure}

By using the concept of an isotropy subgroup,\cite{18}  we found the constraints
for the structural transformations of the system.
In Fig.~\ref{fig:Fig_1}, the B{\"a}rnighausen tree of possible structural relations for \ce{FeAs2} is given,
from where we decided in favor of the following set of allowed group-subgroup pairs:
$Im\bar{3}m \rightarrow I4/mmm \rightarrow Immm \rightarrow Pnnm$.
Based on this prediction,  Fig.~\ref{fig:Fig_2} shows the assembly route as a three-step pathway.
Note that, prior to the first step, we made a tetragonal rebuilding of a single $bcc$ iron lattice,
$Im\bar{3}m \rightarrow I4/mmm$, via equal doubling of the cubic lattice period
$a \rightarrow 2 \times 2.8274$ {\AA} in both longitudinal directions.
This manipulation preformed a superstructure of \ce{Fe} atoms as a starting template qualified for
the incorporation of arsenic dimers. Their initial packing arrangement proceeds by placing them
into a stacked disposition, followed by a small lengthening, $2.8274 \rightarrow 2.8747$ {\AA}, of the lattice
along the tetragonal $c$ direction.
The first step ends in the formation of a putative structure featured by
an orthorhombic space group $Immm$ (labeled as Phase 0).
Because such a trial configuration turns out to be unstable, on the next step, \ce{As} atoms are further packed
to lower the ground state energy.
With respect to the geometry, the subsequent repacking leads to the crossover from the $I$-lattice to the $P$-lattice and
to the reduction in symmetry to the non-isomorphic subgroup $Pnnm$.
Crystal stability is achieved in the coordination environment by rearrangements of the \ce{As} atoms
via the \ce{As-As} bonds  elongation and rotation. This step forms the refined intermediate configuration
("hidden" phase labeled as Phase~I) associated with the conversion of the unfavorable arsenic '$4h$' sites
to '$4g$' ones.
In the final step, the native structure of \ce{FeAs2} (labeled as Phase~II) is formed by
a "square-rectangular" transformation in the planar sublattice of iron atoms.
This transformation generates the difference of the in-plane lattice parameters but keeps the same $Pnnm$ group.

\subsection{Stability and phase competition}
We  performed the mechanical stability tests
for Phases I and II
 to be convinced that the strain energy of a crystal is necessary positive.
The corresponding elastic stability criteria,\cite{kniga} which
make use of the matrices of the elastic constants $C_{ij},$
impose the following restrictions:
 $$C_{11}>0, \, \, \begin{vmatrix}
 C_{11}& C_{12}\\
 C_{21}&C_{22}
\end{vmatrix} \,>\, 0 \, , \,
...\, ,  \, \,  \begin{vmatrix}
 C_{11}& ... & C_{16} \\
 ...   & ... & ... \\
 C_{61}& ... & C_{66}
\end{vmatrix}\,>\,0 \, .$$
Once the matrix elements were determined (as shown in Fig. 3), it was found that
each of the structures represented by Phase~I and Phase~II satisfies the stability criteria.
The results of the PBE+U calculations of the vibrational spectra in harmonic approximation also
validated that  both structures pass the dynamical stability test
(the corresponding frequencies at the $\Gamma$ point are presented in the Appendix).
Furthermore, an analysis of the thermodynamical stability of Phase~I
performed in terms of the heat of formation ($\Delta H_f^0$)
for the theoretical decomposition pathway
$$
\left.\begin{matrix}
\text{FeAs}_\text{2} & \longrightarrow  & \text{Fe} & + & \text{As}-\text{As} \\
 &  & \text{(bcc/solid)} & & \text{(dimer array)}
\end{matrix}\right.
$$
confirmed that  Phase~I is stable with $\Delta H_f^0=\,-178$ kJ/mol.
\begin{figure}[H]
\includegraphics[width=0.55\columnwidth,keepaspectratio=true]{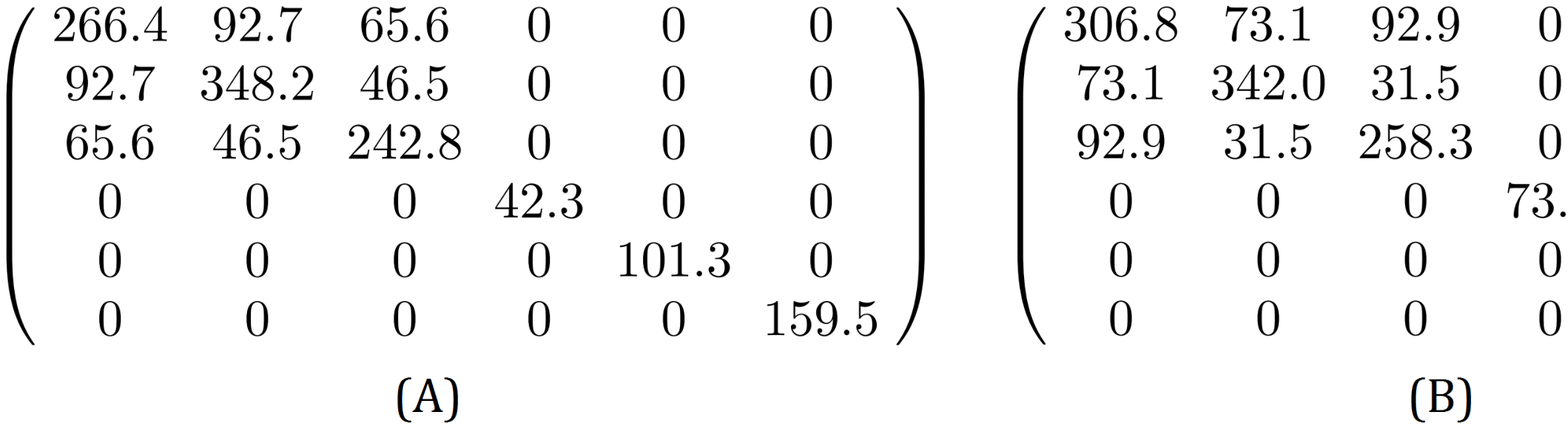}
\caption{The stiffness matrix C for (A) Phase~I, (B) Phase~II, and (C) hypothetical  \ce{FeAs2} polymorph
{\it{Imm2}} symmetry. The nine independent elastic constants C$_{ij}$ in (GPa) were calculated using the PBE+U method.}
\end{figure}

To understand why Phase~I suffers transformation into Phase~II, we first point out
that they differ in total energies by $54$ kJ/mol.
The remarkable fact here is that these phases are closely related stable crystalline modifications of \ce{FeAs2}
and differ mostly by the planar geometry of iron atoms. However, the intermediate Phase~I is less stable
because an enhancement of the planar anisotropy through an in-plane stretching/contraction of the  \ce{Fe-Fe}
distances further lowers  the ground state energy.
In the stereochemical aspect, this result emphasizes that the reason why the stable system undergoes
2D-planar reshaping is very unusual. The energy difference between Phases I and II is large,
whereas the corresponding change of the unit-cell volume is dramatically small ($\sim 0.7$\%).
We note further that the shear ($G_V$) and bulk ($B_V$) moduli estimated in the Voigt approximation\cite{Voigt1928}
are larger than those of Phase~I (Table I).
\begin{table}[htbp]
\caption{\label{tab:Table_data_03}
Shear modulus $G_V$ and bulk modulus $B_V$ in GPa for Phases~I~and~II, estimated in the Voigt approximation\cite{Voigt1928}
from the theoretical elastic constants $C_{ij}$. In the last row, the values of Poisson's ratio ${\nu_V}$ are given.}
\begin{ruledtabular}
\begin{tabular*}{0.5\textwidth}{l c c c}
 & $G_V$ & $B_V$ & ${\nu}_V$ \\
Phase~I & 104.1 & 140.8 & 0.203 \\
Phase~II & 118.8 & 144.7 & 0.178
 \end{tabular*}
  \end{ruledtabular}
\end{table}
This fact,  in terms of the upper bounds for the elastic moduli, represents a clear illustration of
a hardening of the lattice due to the "rectangular" structural reorganization
in the subsystem of the iron atoms.
The corresponding reduction of Poisson's ratio ${\nu}_V$ by ${\sim}12\%$ also indicates the enhancement of
the lattice stability in Phase~II.
Thus, one can conclude that, along with the lack of evidences of any lattice symmetry breaking,
these facts give  direct evidence
that the Phase~I $\rightarrow$ Phase~II transformation is entirely of an electronic origin.
\subsection{Polymorphism of \ce{FeAs2}}
In the context of how the effect of the coordination environment guides the optimization strategy,
we notice that there is another symmetry-allowed group-subgroup pair $Immm \rightarrow Imm2$
(instead of the $Pnnm$ subgroup as shown in Fig.~1) that suggests an alternate version of \ce{As} atoms packing.
A performed theoretical screening predicted the possibility of a novel polymorph of \ce{FeAs2}
with lattice parameters of a=5.425 {\AA}, b=6.279 {\AA}, and c=2.976 {\AA}.
A Table~\ref{tab:Table_data_02} of Appendix gives the set of calculated crystallographic data.
The polymorph has an inverse symmetry broken orthorhombic structure characterized by the polar space group $Im2m$.
To analyze stability issues, we calculated the elastic constants (the stiffness matrix is shown in Figure 3C)
and the harmonic frequencies of the zone centered vibrational modes (the values are displayed in the Appendix).
We found that the $Im2m$ structure fulfills the mechanical and dynamical stability tests.
However, the difference in energy with respect to the native structure is strongly positive, $\Delta E_{tot}=174$ kJ/mol.
Hence, the polymorph of the $Im2m$ symmetry cannot be thermodynamically stabilized under ambient conditions, and
therefore, it is much less feasible than the experimentally observed geometry of Phase~II.

\begin{figure}[H]
\centering
\includegraphics[width=0.45\columnwidth,keepaspectratio=true]{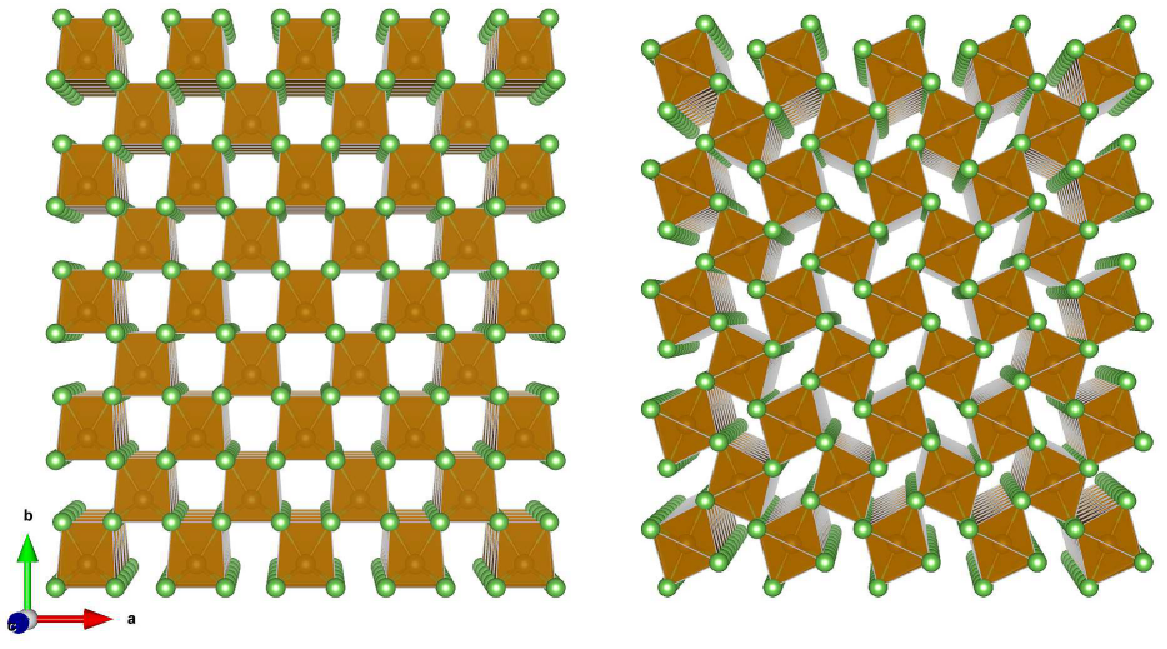}
\caption{\label{fig:Fig_3}
View of the packing scheme of the polymorph and the native phase of \ce{FeAs2} along the  c-axis.
The crystal structures are depicted as a three-dimensional network composed of 1D chains of
corner-sharing \ce{FeAs6} polyhedra.
\ce{FeAs6} polyhedra are shaded in brown. The color codes are the same as in Fig.~2.}
\end{figure}
\begin{figure}[H]
\centering
\includegraphics[width=0.43\columnwidth,keepaspectratio=true]{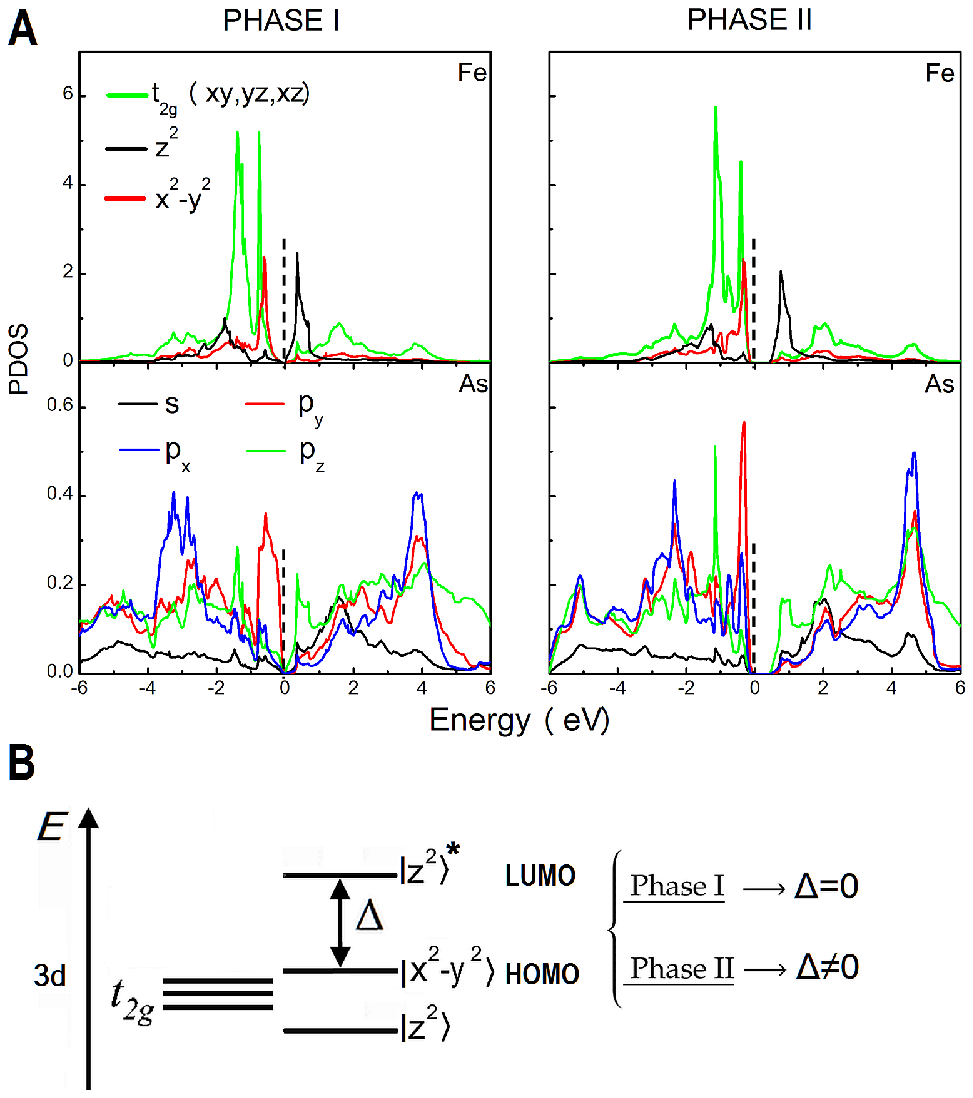}
\caption{\label{fig:Fig_3}
The differences in the electronic structures of Phases~I~and~II obtained from the PBE+U calculations.
(A) Orbital and site projected electron densities of states (PDOS).
Vertical dashed lines indicate the chemical potential position.
(B) An illustrative energy scheme of the $e_g$ doublet, showing the origin of the fundamental gap ($\Delta$)
in the electron spectrum.
Although actual bonding and antibonding orbitals have some hybrid character caused by the
light admixture of the \ce{As} $p$ states (Fig.~\ref{fig:Fig_3}A), frontier \ce{Fe} $e_g$ orbitals
make the major contribution to an orbital character of the gap.}
\end{figure}
Figure 4 illustrates the difference of the arrangement of the  \ce{FeAs6} polyhedra in the crystal structures of
the polymorph and the native phase of \ce{FeAs2}.
In  both geometries, one can observe the void tunnels of keystone and rhomboidal shapes,  the
edges of which are formed by corner-sharing \ce{FeAs6} polyhedra.
Figure 4 also shows that just in the structure of the polymorph,
the tunnels determine the largest interstice space that may present a number of preferential
positions for the incorporation of dopant atoms.
In the context of iron-based materials studies, this interesting fact might
allow us to consider the hypothetical \ce{FeAs2} of $Im2m$ symmetry as
a possible starting compound for a theoretical design of new ternary iron pnictides.
\subsection{Modeling of the structure-property connection}
Figure~\ref{fig:Fig_3} visualizes the respective changes
in the electron subsystem of \ce{FeAs2} in terms of the partial densities of states.
Choosing $bcc$ \ce{Fe} as a reference point and recalling that the assembly is based on
a superstructural architecture derived from the $bcc$ iron lattice, one would expect,
in the context of a $3d$-split configuration, a replication of the effects of atomic-like
\ce{Fe}-$3d$ states in the electron spectrum of \ce{FeAs2}.
It appears to be nearly so, regardless of the $p$-$d$ hybridization effects inspired by
the coordination of \ce{As} ligands.
A comparison shows that the orbital-selective features of \ce{Fe} in \ce{FeAs2},
such as the hybridization character of a $t_{2g}$ manifold of states and stronger localization of
$e_g$ states (Fig.~\ref{fig:Fig_3}A), demonstrate the orbital properties similar to
the $\alpha$-\ce{Fe} electronic structure.\cite{22}
Furthermore, a small pseudogap in the density of the states of Phase~I means that the ligand field strength
is not yet sufficient to overcome  hybridization effects and, hence,
to  completely exclude the system from the metallic range.
However, as shown in Fig.~5, a change of the orbital overlap caused by
the transformation Phase~I $\rightarrow$ Phase~II
leads to the conversion of the pseudogap into a real forbidden gap of $\sim 0.4$ eV.
In contrast to octahedral complexes,
this transformation forms a certain electronic configuration of \ce{Fe} $e_g$ states,
which corresponds to an insulating state of a  orbital nature
(as depicted in Fig.~\ref{fig:Fig_3}B).

To confirm the presented picture of the electron states,
we compared the results of our PBE+U calculations with those obtained
within the G$_{0}$W$_{0}$ approximation.
The G$_{0}$W$_{0}$ calculations resulted in a similar electron structure (see the PDOS dependencies
shown in Fig.~10 of the Appendix), and they reproduced the value of the dielectric gap
very close to that obtained by the PBE+U method.

We can further track how the bonding states of the  $d_{x^{2}-y^{2}}$ symmetry underlie a particular charge ordering
in the orbital sector of \ce{FeAs2}. As schematically illustrated in the upper half of Fig.~6,
the \ce{Fe} atoms occupy the corners of a square with  sides of  length $a$ in Phase~I.
\begin{figure}[H]
\centering
\includegraphics[width=0.30\columnwidth,keepaspectratio=true]{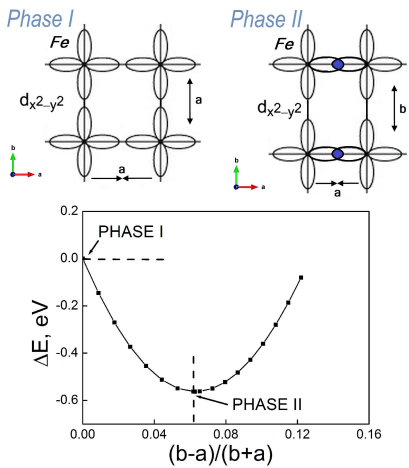}
\caption{\label{fig:Fig_4}
Characteristic patterns describing the  genesis of the nematic ordering  in Phase~II in the orbital sector of
$|x^2-y^2>$ functions that belong to the neighboring sites of the \ce{Fe} sublattice.
The net of equally spaced patterns can be imagined as being formed by a projection of
the $Pnnm$ lattice on the $ab$-plane.
In the bottom half of the figure, the total energy difference with respect to Phase~I vs. the (normalized)
planar deformation is graphed.}
\end{figure}
The orbital overlap generated by the functions $|x^2-y^2>$ remains identical in both planar directions.
Conversely, in Phase~II, where charge redistribution occurs, the full equality of the orbital overlaps
between the $x$ and $y$ directions becomes broken because of the planar tetragonal distortion.
The distortion provides the minimization of the total energy, as shown in the bottom half of Fig.~6.
Based on the results of Refs. \cite{Fernandes, Fernandes1, Yoshizawa, Iye}, this implies
that in the native structure of \ce{FeAs2}, there exists a  electronic ordering of the orbital nematic origin.
The nematicity scenario can be described  as follows: \\
(i) the electron-driven Phase~I $\rightarrow$ Phase~II modification implies the transition to the nematic-like state
in the subsystem of Fe 3$d_{x^{2}-y^{2}}$ valence electron orbitals (so called orbital nematicity \cite{Fernandes, stat}).\\
(ii) in the geometrical context, the ordering is attended with a square $\rightarrow$ rectangular transformation,
with the structural distortion featured by the difference of the planar lattice parameters $(b-a)$; \\
(iii) in the symmetry aspect, because of some freedom provided by the coordination environment effect
within the dedicated orthorhombicity, the transition from Phase~I to Phase~II keeps all the operations of
the space group $Pnnm$ unchanged; \\
and (iv) in terms of the $2D$ internal lattice symmetry related to the $ab$-plane, the distortion can be formally
associated with the additional reduction of the local rotational symmetry from $C_4$ to $C_2$ with respect to
the uniplanar neighboring sites of the \ce{Fe} sublattice.
In addition, one can also emphasize that the charge aspect of the orbital nematicity is that a spatial arrangement of
the \ce{Fe} $3d_{x^{2}-y^{2}}$ orbitals induces charge ordering that can be characterized by the difference of
the overlap integrals between neighboring Fe $d_{x^{2}-y^{2}}$ states along the $x$- and $y$- planar directions.
\subsection{Features of the bonding architecture: an interplay of the ordering and
dynamic equilibration}
\begin{figure}[H]
\centering
\includegraphics[width=0.55\columnwidth,keepaspectratio=true]{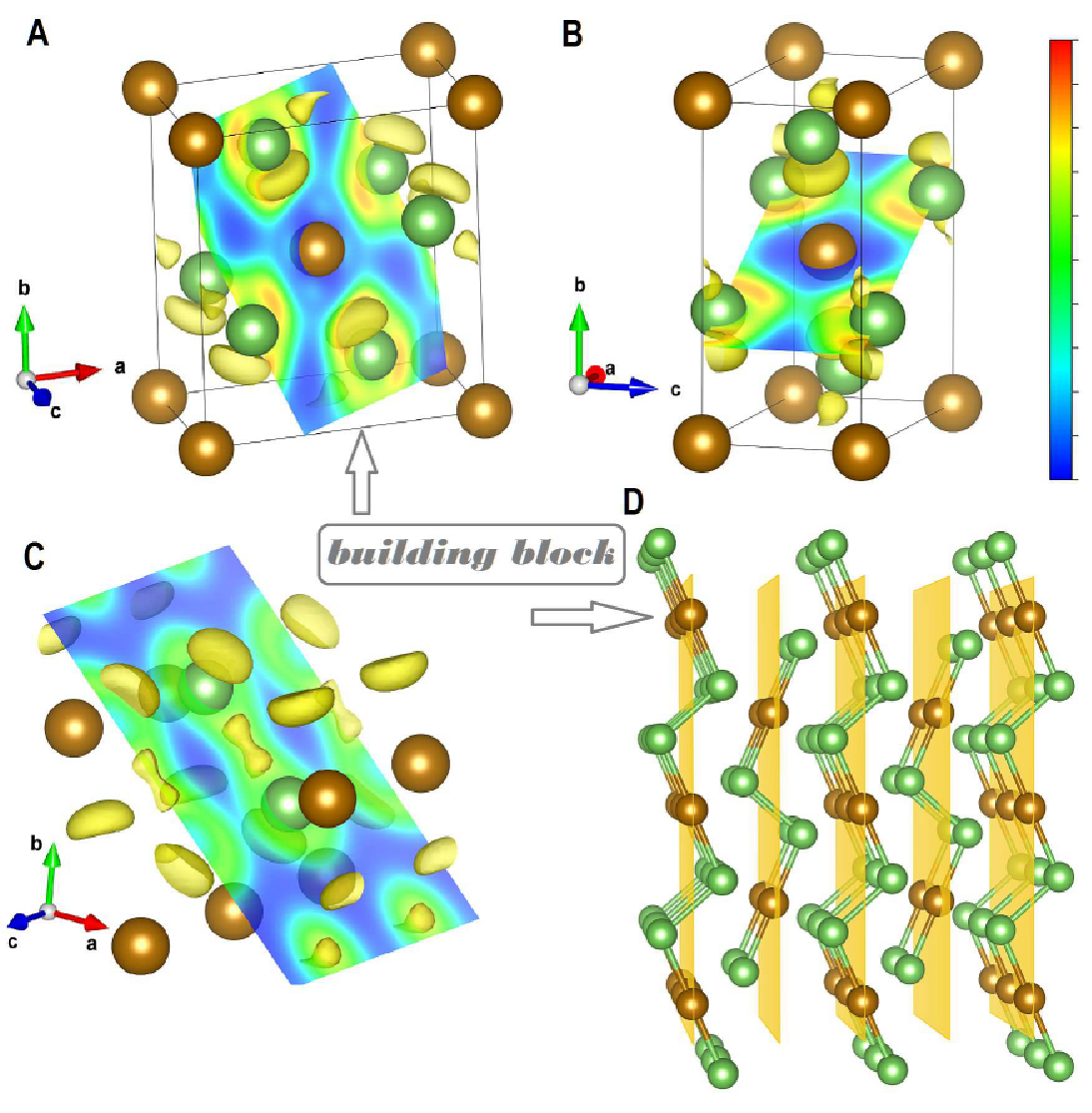}
\caption{\label{fig:Fig_5}
Valence ELF distribution patterns with isosurfaces projected onto the
(A) ($1.5$ $0.75$ $1.2$), (B) ($-1.6$ $3.6$ $0$), and (C)  ($0.68$ $0.88$ $0$) planes
of the \ce{FeAs2} crystal structure. The visualizations were evaluated at ELF$=0.661$.
The numerical values of the total charge partitioning, normalized between ELF$=0$ and ELF$=0.8,$
are illustrated by the color scale between dark blue and red, respectively.
The  lobe-like isosurface (in yellow) displayed near the \ce{As} atoms corresponds to
a lone pair domain.
(D) Schematic model view of the \ce{As-Fe-As} building blocks in the \ce{FeAs2} lattice;
the blocks possess a linear ($D_{{\infty}h}$) geometry and consist of a central \ce{Fe^{2+}}
and two geminal \ce{As^{-}}. The color codes of atoms are the same as in Fig.~\ref{fig:Fig_2}.
}
\end{figure}

To further highlight the effect of coordination environment, 
we point out that the relevant $\sigma$- and $\pi$-interactions driving the covalent binding of ligands to the TM
form an inclined disposition of \ce{As-Fe-As} building blocks in the lattice.
The corresponding structural motifs are shown in Fig.~\ref{fig:Fig_5}D.
From the aspect of dynamics  the geometry of the \ce{As-Fe-As} building block promotes  promotes the interplay between interatomic motions
and valence electronic states.
To explore how this interplay arises, note that, in the relatively low-symmetry insulators,
possessing a high electronic polarizability, small variations in the spatial positions of atoms,
such as vibrations, may alter the static bonding topology by involving additional
displacement-induced symmetry-allowed combinations of the frontier TM $d$ and ligand $p$ orbitals
overlapping (dynamic $p$-$d$ hybridization).\cite{26,27}
The results of our tests showed that dynamic redistribution of the valence charge density
has a pronounced effect in \ce{FeAs2}. The evidence can be seen in Table~\ref{tab:Table_eps}
and Figures 7A-7C and 8.
\begin{table}[htbp]
\caption{\label{tab:Table_eps}
Diagonal elements of the matrix of the Born effective charges (in $|e|$) along with the
theoretical values of the macroscopic dielectric constant $\epsilon_{\infty}$.
}
\begin{ruledtabular}
\begin{tabular}{lccc}
  & $x$ & $y$ & $z$ \\
    \hline
\ce{Fe} & $-5.82$ & $-5.05$ & $-6.94$ \\
\ce{As} & $+2.91$ & $+2.52$ & $+3.49$ \\
\hline
 $\epsilon_{\infty}$ & $27.0$ & $33.0$ & $24.4$ \\
  \end{tabular}
  \end{ruledtabular}
\end{table}
\begin{figure}[H]
\centering
\includegraphics[width=0.30\columnwidth,keepaspectratio=true]{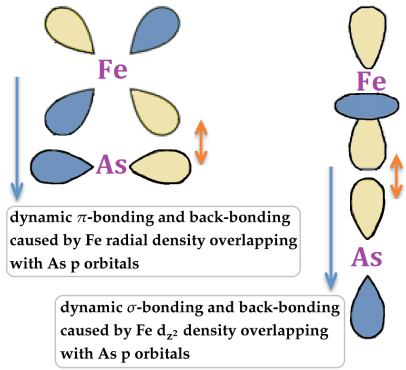}
\caption{\label{fig:Fig_6}
Schematic diagrams illustrating the dynamic dative nature of synergic bonds
in terms of the $\pi$- and $\sigma$- components.
}
\end{figure}

The first observation is that the Born effective charges of Table~\ref{tab:Table_eps}
are enormously anomalous as compared with the formal ionic charges of the ligand-field model.
The second observation is based on the analysis of ELF.
It displays the peculiar role that $p$-type lone pairs play at the vibrational
contact.
These $s^{2}p^{4}$  pairs are localized and ordered around
the geminal \ce{As^{-}}.
The stereochemical activity of the pair  mostly has a dynamical character,
i.e., it may benefit the local delocalization through the dynamical back bonding
from the ligand lone-pair to the central \ce{Fe^{2+}} $3d$ states.
The "virtual" transfer of the lone pair promotes a  iron-reductive (arsenic oxidative)
coupling that brings them an enhanced negative (positive) effective charge (Table~\ref{tab:Table_eps}).
Interestingly, the similar effect of "virtual" presence of a lone pair through electron-phonon coupling
was considered for the  \ce{Te}-II phase.\cite{28}
Because the ligands stay coordinated to \ce{Fe} at the vibrational contact,
the reverse process is the donation $3d$(\ce{Fe}) $\rightarrow$ $4p$(\ce{As}) to restore
the former localization of the lone-pair.
Both processes driven by proper IR-active optical vibrations are therefore dynamically strung together
by the underlying (static) bonding architecture.
Thus, the strong symmetry-allowed vibrational contacts of the $p$ and $d$ orbitals mediate in \ce{FeAs2}
the excess negative charge-flow from the ligands onto the metal and back.
This can be referred to as the dynamic component of the net charge-transfer, extra-large in magnitude
but opposite in sign.
Moreover, in the sense of dynamics, the back-bonding phenomenon correlates well with the fact that
the Bader effective charges calculated for \ce{Fe} and \ce{As} have almost zero indicative values.
This result simply indicates that, in the sector of the dominating covalency,  "external" forcing
such as the accompanying vibrations can be responsible for the additional charge asymmetry.

In the context of such an effect of dynamic equilibration, it is interesting to note
that a similar situation characterized by the importance of polarization effects
in the direct interaction of ligands lone pairs with the metal exists in Zn(II) complexes.\cite{29}
\section{Conclusion}
In the present work, we performed a DFT-based analysis of the structural chemistry
of the binary pnictide \ce{FeAs2}. We showed how  the electronic features
caused by the  behavior of the $3d$ electrons of a TM can be understood in terms of the
investigation of  the structural trends driven by TM-ligand interactions and the synthetic construction
of a preferred architecture accompanied by using \ce{Fe-Fe} bonds.
Starting from a known chemical composition and having made the proper choice of symmetry constraints
as an integral part of the assembling strategy, we  reproduced the native architecture of the loellingite
from first-principles. Wr also  considered, for the first time, two original electronic properties of \ce{FeAs2},
a nematic-like ordering in the subsystem of Fe 3$d_{x^{2}-y^{2}}$ valence electron orbitals
and an extra-large dynamic charge-transfer caused by a  lone-pair configuration.
The ordering results from the design-imposed structural adjustment of iron planar positions,
whereas the negative dynamic charge-transfer employs the dynamical lability of the \ce{Fe-As} interactions
in the covalent framework.
The nematic ordering represents a macroscopic cooperative phenomenon in the electron subsystem of \ce{FeAs2}.
In a sense of crystal chemistry, this is a manifestation of the structure-property relationship
because the  electronic topological effect is directly driven by the lattice degrees of freedom
via the difference of the planar arrangements of the \ce{Fe} atoms.

The interplay between the metallicity and covalence, which is a constructional feature of
the \ce{FeAs2} crystal design, may be responsible for a certain softness of the \ce{Fe-As} covalent bonds.
In this context, some interesting parallels with the general design principles of
dynamic covalent chemistry\cite{30} could be suggested.
A tetragonally distorted form of $bcc$ iron involved in the three-step formation pattern of
\ce{FeAs2} (Fig.~\ref{fig:Fig_2}) serves as a predesigned bulk superstructure that functions
as a template to distribute ligands in the condensed phase.
Preferential binding between the \ce{Fe} centers and \ce{As} ligands,
which corresponds to the so-called "covalent capture",
makes the structural motif of the template a basic part of all possible structures (including the native one)
contained in the formation route.
The distribution proceeds through a sequential "proof-reading" generation of two intermediate geometries
because the metallic bonding of the template scales formation of covalent bonds by searching for the global minimum.
The last structural correction takes place largely inside of the template and deals with the utilization
of \ce{Fe-Fe} bonds  to provide the state with a nematic ordering.
The coordination environment effect is that the corresponding coordination and stabilization processes
place \ce{As} atoms into  favorable positions.
On the other hand, the covalent capture maintains the usual labile character of \ce{Fe-As} ligand interactions.
Thus, the coupling of two effects, the coordination environment and dynamic equilibration,
through the assembly of the most dynamically stable structure of \ce{FeAs2} is another interesting fact
in the context of the close interplay of the lattice and the electron degrees of freedom in this material.
\begin{acknowledgments}
The present work was supported by the European Regional Development Fund
(Centre of Excellence "Mesosystems: Theory and Applications", TK114)
and the ESF Grant No. 8991.
The authors would like to thank Dr. M. Klopov for attention to this work.
\end{acknowledgments}
\section{Appendices}
\appendix*
\section{Crystallographic data}
\vspace{-0.7cm}
\begin{table*}[htbp]
\caption{\label{tab:Table_data_01}
Comparison of the relaxed equilibrium geometries (orthorhombic unit-cells, volumes,
atomic positions, interatomic distances, and angles).
The experimental data are given in the last column.
A structure of symmetry $Immm$ (Phase 0) suffers a loss of stability.
At the symmetry break, which corresponds to $Immm\,\rightarrow \,Pnnm$ (Phase 0 $\rightarrow$ Phase~I),
the lattice constants do not change.
The orthorhombic symmetry of Phases 0 and I differs by  sets of \ce{As} coordinates.
The differences between the phases are illustrated in Fig.~\ref{fig:Fig_A1}.
}
\begin{ruledtabular}
\begin{tabular}{lcccc}
\ce{FeAs2}  & Phase 0 & Phase~I & Phase~II & Expt.\cite{2} \\
            & Immm    & Pnnm    & Pnnm     & Pnnm \\
    \hline
$a$({\AA})$=$ & $5.6548$ & $5.6548$ & $5.2902$ & $5.3012(6)$ \\
$b$({\AA})$=$ & $5.6548$ & $5.6548$ & $5.9994$ & $5.9858(5)$ \\
$c$({\AA})$=$ & $2.8747$ & $2.8747$ & $2.8747$ & $2.8822(4)$ \\
$V$({\AA}$^3$)$=$ & $91.92$ & $91.92$ & $91.24$ & $91.46$ \\
    \hline
Wyckoff positions &  &  &  &  \\
\ce{Fe} & $2a$ $0\,|\,0\,|\,0$ & $2a$ $0\,|\,0\,|\,0$ & $2a$ $0\,|\,0\,|\,0$ & $2a$ $0\,|\,0\,|\,0$ \\
\ce{As} & $4h$ $0\,|\,0.2967\,|\,0.5$ & $4g$ $0.1838\,|\,0.3654\,|\,0$ & $4g$ $0.1790\,|\,0.3616\,|\,0$ & $4g$ $0.1763(10)\,|\,0.3624(7)\,|\,0$ \\
    \hline
nearest neighbor        &  &  &  &  \\
bond lengths (in {\AA}) &  &  &  &  \\
\ce{As-Fe}              & $2.209$ & $2.417$; $2.313$ & $2.375$; $2.367$ & $2.388(4)$; $2.362(4)$ \\
\ce{As-As}              & $2.299$ & $2.576$ & $2.519$ & $2.492(7)$ \\
\ce{Fe-Fe}              & $2.875$ & $2.875$ & $2.875$ & $2.882$ \\
 bond angles (in grad.)   &  &  &  &  \\
\ce{Fe-As-Fe}           & $81.2$; $135.8$ & $73.0$; $127.9$ & $74.5$; $127.3$ & $74.3(1)$; $127.0(1)$ \\
\ce{As-Fe-As}           & $81.2$; $135.8$ & $107.0$; $87.1$; $92.9$ & $105.5$; $92.0$; $88.0$ & $105.7(1)$; $91.9(2)$; $88.1(2)$ \\
  \end{tabular}
  \end{ruledtabular}
\end{table*}
\vspace{-0.2cm}
\begin{figure}[H]
\centering
\includegraphics[width=0.8\columnwidth,keepaspectratio=true]{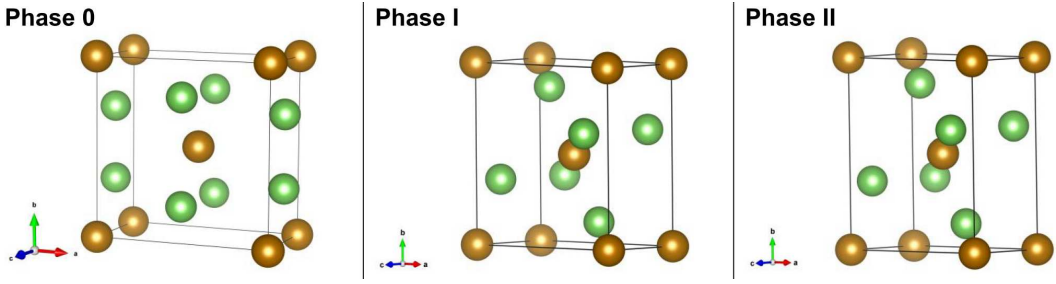}
\caption{\label{fig:Fig_A1}
Schematic presentation of the unit cells.
The color codes of atoms are the same as in Fig.~\ref{fig:Fig_2}.}
\end{figure}

\begin{table}[htbp]
\caption{\label{tab:Table_data_02}
Relaxed equilibrium structure of a predicted polymorph of \ce{FeAs2}
featured by  space group No $44$.
Schematic presentation of an unit cell is shown in Fig.~\ref{fig:Fig_B1}.
}
\begin{ruledtabular}
\begin{tabular}{ll}
Crystal system & Orthorhombic\\
Space group & Im2m ($44:bca$ choice) \\
$a$({\AA})$=$ & $5.425$ \\
$b$({\AA})$=$ & $6.279$ \\
$c$({\AA})$=$ & $2.976$ \\
$Z$ & $2$ \\
\hline
Wyckoff positions &  \\
\ce{Fe} & $2a$ $0\,|\,0\,|\,0$ \\
\ce{As} & $4d$ $0.2290\,|\,0.3317\,|\,0$  \\
 \end{tabular}
  \end{ruledtabular}
\end{table}
\begin{figure}[H]
\centering
\includegraphics[width=0.25\columnwidth,keepaspectratio=true]{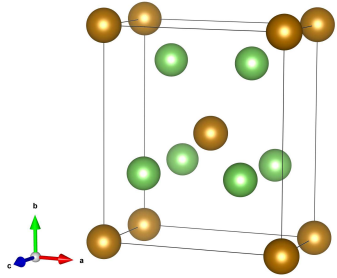}
\caption{\label{fig:Fig_B1}
Schematic presentation of a unit cell of the predicted \ce{FeAs2} polymorph
with a crystallographically distinct equilibrium arrangement of \ce{As} atoms.
The color codes of atoms are the same as in Fig.~\ref{fig:Fig_2}. }
\end{figure}
\section{Dynamical stability test}

List of zone centered vibrational modes calculated in the harmonic approximation (in cm-1):
\\



\begin{tabular}{lrrlllllllllllll}
Phase~I         & 109& 130& 130& 135& 189& 190& 204& 211& 227& 238& 254& 258& 280& 307& 325\\
Phase~II        & 126& 137& 144& 157& 181& 229& 230& 231& 238& 259& 267& 271& 273& 305& 330 \\
Imm2 polymorph  &  36&  93& 136& 138& 141& 163& 171& 197& 215& 248& 257& 262& 268& 271& 315 \\
 \end{tabular}

\section{Electron structure of the Phase~II in G$_0$W$_0$ approach}
\begin{figure}[H]
\centering
\includegraphics[width=0.5\columnwidth,keepaspectratio=true]{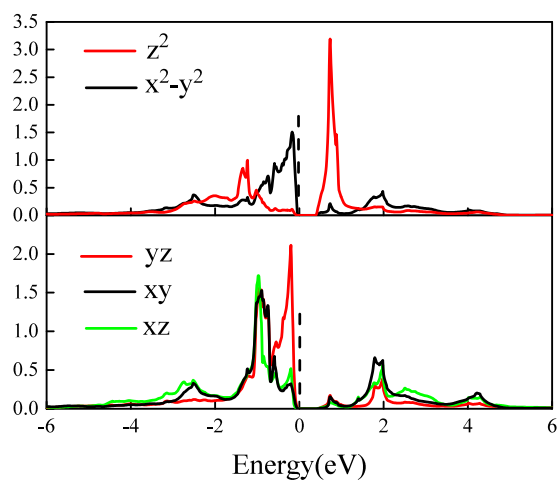}
\caption{The partial densities of d-states of Fe in the Phase~II obtained
by using G$_0$W$_0$ method.
Vertical dashed lines indicate the chemical potential position.
}
\end{figure}

\bibliographystyle{apsrev}

\begin{thebibliography}{10}
\bibitem{stat} Different aspects and scenarios of the electronic nematicity in superconducting materials based on iron pnictides
and its role in the context of the interplay of structure, charge and magnetic degrees of freedom are currently the subject of
 intensive research work (e.g.,\cite{Fernandes, Fernandes1, Yoshizawa, Iye, Hoffman}).
\end{thebibliography}

\end{document}